\documentclass[epj]{webofc}
\usepackage[utf8]{inputenc}
\usepackage[varg]{txfonts}   % Web of Conferences font
\usepackage{booktabs}
\usepackage{xcolor}
\usepackage{feynmf}

\definecolor{darkred}{rgb}{0.4,0.0,0.0}
\definecolor{darkgreen}{rgb}{0.0,0.4,0.0}
\definecolor{darkblue}{rgb}{0.0,0.0,0.4}
\usepackage[bookmarks,linktocpage,colorlinks,
    linkcolor = darkred,
    urlcolor  = darkblue,
    citecolor = darkgreen]{hyperref}
\usepackage{subfigure}
\usepackage{braket}
\usepackage{amsmath}
\usepackage{graphicx}
\usepackage{slashed}
\wocname{EPJ Web of Conferences}
\woctitle{Lattice2017}
\newcommand{\matel}[2]{\braket{\bar{#2}|#1|#2}}
\newcommand{\mateldiff}[3]{\braket{#1|#2|#3}}
\newcommand{\glow}[1]{\gamma_{#1}}

\begin{document}
%%%%%%%%%%%%%%%%%%%%%%%%%%%%%%%%%%%%%%%%%%%%%%%%%%%%%%%%%%%%%%%%%%%%%%%%%%%%%
%
\selectlanguage{english}
%----------------------------------------------------------------------------
\title{%
BSM Kaon Mixing at the Physical Point
}
%----------------------------------------------------------------------------
\author{%
\firstname{Peter} \lastname{Boyle}\inst{1,3}\and
\firstname{Nicolas} \lastname{Garron}\inst{2,3} \and
\firstname{Julia}  \lastname{Kettle}\inst{1,3}\fnsep\thanks{Speaker, \email{J.R.Kettle-2@sms.ed.ac.uk} }
\firstname{Ava} \lastname{Khamseh}\inst{1,3} \and
\firstname{Justus Tobias} \lastname{Tsang}\inst{1,3} 
% etc.
}
%----------------------------------------------------------------------------
\institute{%
University of Edinburgh
\and
University of Liverpool
\and
RBC-UKQCD
}
%----------------------------------------------------------------------------
\abstract{%

We present a progress update on the calculation of beyond the standard model (BSM) kaon mixing matrix elements at the physical point. Simulations are performed using 2+1 flavour domain wall lattice QCD with the Iwasaki gauge action at 3 lattice spacings and with pion masses ranging from 430 MeV to the physical pion mass. 

}
%----------------------------------------------------------------------------
\maketitle
%----------------------------------------------------------------------------
\section{Introduction}\label{intro}

\begin{figure}[h]	
	\begin{center}
		\begin{fmffile}{KaonMixing} 
				\begin{fmfgraph*}(100,45) 
					\fmfleft{i1,i2} 
					\fmfright{o1,o2} 
					\fmflabel{$\bar{d}$}{i2} 
					\fmflabel{$d$}{o1} 
					\fmflabel{$s$}{i1} 
					\fmflabel{$\bar{s}$}{o2} 
					\fmf{fermion}{i1,v1} 
					\fmf{fermion,tension=.5,label=$\bar{u},,\bar{c},,\bar{t}$,1.side=right}{v1,v3} 
					\fmf{fermion}{v3,o1} 
					\fmf{fermion}{o2,v4} 
					\fmf{fermion,tension=.5,label=$\bar{u},,\bar{c},,\bar{t}$,1.side=right}{v4,v2} 
					\fmf{fermion}{v2,i2} 
					\fmf{photon,tension=0,label=$W$,1.side=left}{v1,v2} 
					\fmf{photon,tension=0,label=$W$,1.side=left}{v3,v4} 
					\fmfdotn{v}{4} 
				\end{fmfgraph*}     
		\end{fmffile} 
	\caption{W exchange box diagram}
	\label{fig:kkmixing}
	\end{center}
\end{figure}
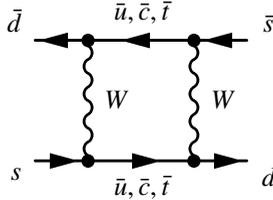

Kaon mixing is a flavour changing neutral current process in which a neutral kaon ${K^0}$ oscillates with its anti-particle $\bar{K^0}$. In the standard model (SM) it is dominated by box diagrams such as the one shown in figure \ref{fig:kkmixing}.

\vspace{-0.9cm}

We separate out the long distance contributions, using the operator product expansion (OPE), into a matrix element $\matel{O_1}{K^0}$ where $O_1$ is the SM four quark operator shown in equation \ref{eq:opbasis}. It has (vector-axial)$\times$(vector-axial) Dirac structure in the SM as a result of the W vertices. Beyond the SM, when the mediating particle is not constrained by the standard model flavour changing vertices, effective operators with other Dirac structures are possible. We can construct a full (SUSY) basis \cite{Gabbiani:1996hi} of five parity-even four-quark operators:

\begin{equation}
\begin{split}
%\begin{align}
O_1 &=(\bar{s}_a \glow{\mu}(1-\glow{5})d_a)(\bar{s}_b \glow{\mu}(1-\glow{5})d_b) \\
O_2 &= (\bar{s}_a (1-\glow{5})d_a)(\bar{s}_b (1-\glow{5})d_b)\\
O_3 &= (\bar{s}_a (1-\glow{5})d_b)(\bar{s}_b (1-\glow{5})d_a)\\
O_4 &= (\bar{s}_a (1+\glow{5})d_a)(\bar{s}_b (1+\glow{5})d_b)\\
O_5 &= (\bar{s}_a (1+\glow{5})d_b)(\bar{s}_b (1+\glow{5})d_a),\\
\label{eq:opbasis}
\end{split}
%\end{align}
\end{equation}
which appear in the effective $\Delta S = 2$ Hamiltonian as
\begin{equation}
\mathcal{H}^{\Delta S = 2} = \sum_{i=1}^5 C_i(\mu) O_i(\mu).
\end{equation}

Whilst the Wilson coefficients $C_i(\mu)$ depend on the physics of the particular BSM model studied, the operators themselves are model independent.

\subsection{Motivation}

The BSM matrix elements have not been as widely studied as the standard model $B_k$. There have been calculations by RBC-UKQCD \cite{Boyle:2012qb}\cite{Garron:2016mva}, ETM \cite{Bertone:2012cu}\cite{Carrasco:2015pra} and SWME \cite{Bae:2013tca}\cite{Jang:2014aea}\cite{Jang:2015sla}, but there are some tensions between the results from different collaborations. These differences can be seen in table \ref{tab:compBSM} and are summarised in the most recent FLAG report \cite{Aoki:2016frl}. 

The most recent BSM kaon mixing study by RBC-UKQCD \cite{Garron:2016mva} sought to address these tensions and proposed that they arose from different choices in the renormalisation methods applied. It was argued that the new RI-SMOM scheme introduced was better behaved than the more commonly used RI-MOM scheme. 
This work aims to improve upon the precision of those results by including a third lattice spacing and ensembles with physical pions. By obtaining more precise results we should be able to comment on the renormalisation scheme's role in the obeserved tension.

\begin{table}
	\caption{Results from calculations of the BSM bag parameters, presented in $\overline{\text{MS}}(\mu=3\textrm{GeV})$, from RBC-UKQCD, SWME and ETM are shown here. We can see that for $B_4$ and $B_5$ there are tensions between the different calculations. RBC-UKQCD results renormalised via the intermdiate RI-MOM scheme agree with the ETM results renormalised via the same scheme. RBC-UKQCD results obtained via RI-SMOM, were obtained via two different schemes, ($\glow{\mu},\slashed{q})$ as detailed in \cite{Boyle:2017jwu}, which agree with each other and with the SWME calculations which use a 1 loop intermediate scheme. This would suggest that the difference in intermediate schemes is responsible for the tensions exhibited, and we conjecture the exceptional infrared behaviour of the RI-MOM scheme, that
		was tamed by modelling pion poles, is the most likely source. }
	\label{tab:compBSM}
	\resizebox{\textwidth}{!}{
	\begin{tabular}{ c | c c | c | c | c c }
		\hline \hline
		& ETM12\cite{Bertone:2012cu} & ETM15\cite{Carrasco:2015pra} & RBC-UKQCD12\cite{Boyle:2012qb} &  SWME15\cite{Jang:2015sla}& \multicolumn{2}{c}{RBC-UKQCD16\cite{Garron:2016mva}}  \\
		\hline \hline
		$n_f$ & 2 & 2+1+1 & 2+1 & 2+1  & 2+1 & 2+1 \\
		scheme & RI-MOM & RI-MOM & RI-MOM & 1 loop & RI-SMOM & RI-MOM \\
		\hline
		$B_2$ & 0.47(2) & 0.46(3)(1) & 0.43(5) & 0.525(1)(23) & 0.488(7)(17) & 0.417(6)(2) \\
		$B_3$ & 0.78(4) & 0.79(5)(1) & 0.75(9) & 0.773(6)(35) & 0.743(14)(65) & 0.655(12)(44) \\
		$B_4$ & 0.76(3) & 0.78(4)(3) & 0.69(7) & 0.981(3)(62) & 0.920(12)(16) & 0.745(9)(28) \\
		$B_5$ & 0.58(3) & 0.49(4)(1) & 0.47(6) & 0.751(7)(68) & 0.707(8)(44) & 0.555(6)(53) \\
		\hline \hline
	\end{tabular} 
}
\end{table}

%----------------------------------------------------------------------------------------
\section{Parameterisation of the Matrix Elements}

\subsection{Bag Parameters}

The renormalised bag parameter is defined as the ratio of the matrix element over its vacuum saturation approximation value,

\begin{equation}
\label{eq:BagGen}
B_i(\mu) = \frac{ \matel{O_i(\mu)}{K^0} }{\matel{O_i(\mu)}{K^0}}_{\textrm{VSA}.}
\end{equation}

At leading order, the forms of the SM and BSM bag parameters are given by,

\begin{minipage}[b]{0.425\linewidth}
\begin{equation}
\label{eq:BagSM}
B_1(\mu)= \frac{ \matel{O_1(\mu)}{K^0} }{\frac{8}{3}m_K^2 f_K^2},
\end{equation}

\end{minipage}
\begin{minipage}[b]{0.525\linewidth}
\begin{equation} 
\label{eq:BagBSM}
B_i(\mu) = \frac{ (m_s(\mu) + m_d(\mu))^2 }{N_i m_K^2 f_K^4}   \matel{O_i(\mu)}{K^0}.
\end{equation}
\end{minipage}

\vspace{0.25cm}
The factors $N_i$ depend upon the basis in which we're working. As we work in the SUSY basis, $N_i = (\frac{8}{3},\frac{-5}{3},\frac{1}{3},2,\frac{2}{3})$.

\subsection{Ratio Parameters}

Ratio parameters, $R_i$, are another parametrisation of the BSM matrix elements. The idea of using ratios to define parameters was proposed in \cite{Donini:1999nn}, following the forms given in  \cite{Babich:2006bh}, we define the ratio parameters as,

\begin{equation}
R_i\bigg( \frac{m^2_P}{f^2_P} , a^2, \mu\bigg) = 
\bigg[ \frac{f_K^2}{m_K^2} \bigg ]_{Exp.} 
\bigg[ \frac{m_P^2}{f_P^2} 
\frac{\matel{O_i(\mu)}{P}}{ \matel{O_1(\mu)}{P} } \bigg]_{Lat.,} 
\end{equation}

where $P$ denotes the simulated strange-light pseudoscalar meson.
At the physical point they reduce to direct ratios of the BSM to SM matrix elements ,

\begin{equation}
R_i(\mu) = R_i\bigg( \frac{m^2_K}{f^2_K} , a=0, \mu\bigg) = 
\frac{ \matel{O_i(\mu)}{K^0} }{  \matel{O_1(\mu)}{K^0} }  .
\label{eq:Ratio-phys}
\end{equation}

The ratio parameters have some advantages over the bag parameters. As there is no explicit dependence on the quark masses, the matrix elements can be recovered from the ratio parameters $R_i$, the SM bag parameter $B_K$ and the experimentally measured kaon mass and decay constant alone. In addition we can expect some cancellation of errors due to the similarity of the numerator and denominator.

%----------------------------------------------------------------------------
\section{Lattice Implementation}\label{sec-1}

We use RBC-UKQCD's $ n_f=2+1 $ gauge ensembles generated with the Iwasaki gauge action \cite{Iwasaki:1985we}\cite{Okamoto:1999hi}. Our ensembles have a DWF action with either the M\"obius \cite{Brower:2004xi} or Shamir \cite{Shamir:1993zy} kernel. These ensembles span 3 lattice spacings; (C)oarse, (M)edium, and (F)ine. C0 and M0 have physical pion masses and all ensembles have physical valence strange quark masses. The details of the ensembles are shown in table \ref{tab:enspar}. The ensembles C0 and M0 have been described in more detail in \cite{Blum:2014tka}, and F1 in \cite{Boyle:2017jwu}.
\begin{table}[h]
	\centering

	\caption{The main parameters of the ensembles included in our analysis are summarised here. C, M and F stand for coarse, medium and fine, respectively. M and S stand for Moebius and Shamir kernels respectively. The propagators had either Z2 wall (Z2W) or  Z2 Gaussian Wall (Z2GW) sources with the latter including source smearing. }
	\label{tab:enspar}
		\resizebox{\textwidth}{!}{
	\begin{tabular}{c | c c | c c c c c | c c c c}
		\hline
		\hline
		name & $L/a$ & $T/a$ & kernel & source & $a^{-1}[\textrm{GeV}]$ & $m_\pi[\textrm{MeV}]$ & $n_{configs}$ & $am_l^{uni} $  & $am_s^{sea}$ & $am_s^{val}$ & $am_s^{phys}$  \\
		\hline
		C0 & $48$ & $96$ & M & Z2GW & 1.7295(38) & 139 & 90 & 0.00078 & 0.0362 & 0.0358 &  0.03580(16) \\
		C1 & $24$ & $64$ & S & Z2W & 1.7848(50) & 340 & 100 & 0.005 & 0.04 & 0.03224 & 0.03224(18) \\
		C2 & $24$ & $64$ &  S & Z2W & 1.7848(50) & 430 & 101 & 0.01 & 0.04 & 0.03224 & 0.03224(18) \\
		\hline
		M0 & $64$ & $128$ & M & Z2GW & 2.3586(70) & 139 & 82 & 0.000678& 0.02661 & 0.0254 &0.02539(17) \\
		M1 & $32$ & $64$ & S & Z2W & 2.3833(86) & 303 & 83 & 0.004 & 0.02477 & 0.02477 & 0.02477(18) \\
		M2 & $32$ & $64$ & S & Z2W & 2.3833(86) & 360 & 76 & 0.006 & 0.02477 & 0.02477 & 0.02477(18) \\
		\hline
		F1 & $48$ & $96$ & M & Z2W & 2.774(10) & 234 & 82 & 0.002144 & 0.02144 & 0.02144 & 0.02132(17)  \\
		\hline
		\hline
	\end{tabular}
}
\end{table}

%----------------------------------------------------------------------------------------
\subsection{Correlator Fitting}\label{sec-2}

We define the two-point and three-point functions as

\begin{minipage}[t]{0.475\linewidth}
\begin{equation}
c_{\mathcal{O}_1 \mathcal{O}_2}(t_i,t) = \sum_x \braket{\mathcal{O}_2(x,t)\mathcal{O}_1(x_i,t_i)},
\label{eq:2pt}
\end{equation}
\end{minipage}
\begin{minipage}[t]{0.475\linewidth}
\begin{equation}
c_{O_k}(t_i,t,t_f)
= \mateldiff{P(t_f)}{O_k(t)}{P(t_i)},
\label{eq:3pt}
\end{equation}
\end{minipage}

where $\mathcal{O}_{1/2}$ denote bilinear operators which in this work are either $\mathbb{P}$, the pseudo-scalar density, or $\mathbb{A}_0$, the temporal component of the local axial current and $O_k$ are the four-quark operators.

At large times the ground state dominates and we can fit the two-point correlators to a $\cosh$ or $\sinh$ function (depending on the Dirac structure of $\mathcal{O}_1$ and $\mathcal{O}_2$) to measure the pseudoscalar masses and amplitudes.

\begin{equation}
c_{\mathcal{O}_1 \mathcal{O}_2}(t_i,t) \xrightarrow[t_i \ll t \ll T]{} \frac{a^4\braket{0|\mathcal{O}_2|P}\braket{P|\mathcal{O}_1|0}}{2am_P}
\bigg(e^{-m_P(t-t_i)}\pm e^{-m_P(T-(t-t_i))}\bigg)
\label{eq:2ptfit}
\end{equation}

Taking ratios of the correlators to measure $B_i^{\textrm{Lat}}$ and $R^{\textrm{Lat}}_i$, as shown in equations \ref{eq:Rcorrs} and \ref{eq:Bcorrs}, they plateau far from the lattice time extent boundaries and we can fit to a constant. 
\begin{equation}
R_k^{\textrm{Lat}}(t_f,t,t_i) = \frac{c_{O_k}(t_i,t,t_f)}{c_{O_1}(t_i,t,t_f)}\xrightarrow[t_i \ll t  \ll t_f \ll T]{} \frac{\matel{O_k}{P}}{\matel{O_1}{P}}
\label{eq:Rcorrs}
\end{equation}
\begin{equation}
B_k^{\textrm{Lat}}(t_f,t,t_i) = \frac{1}{N_k} \frac{ c_{O_k}(t_i,t,t_f) }
{ c_{\bar{P}P}(t_i,t) c_{P\bar{P}}(t_i,t)}
\xrightarrow[t_i \ll t  \ll t_f \ll T]{} \frac{1}{N_k}\frac{\matel{O_k}{P}}{\braket{\bar{P}|\mathbb{P}|0} \braket{0|\mathbb{P}|P}}, \; \; \; \; k>1.
\label{eq:Bcorrs}
\end{equation}

\begin{figure}
	\centering
	\subfigure[The fit of $B_2^{\textrm{Lat}}$ measured on C0 to a constant.]{ \includegraphics[width=0.48\linewidth]{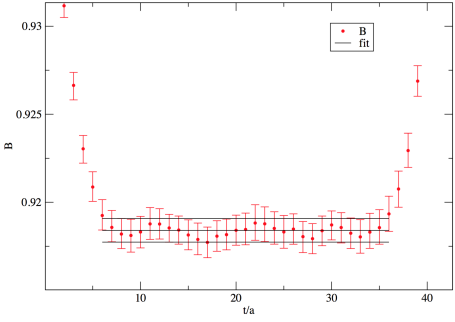} }
	\subfigure[$R^{\textrm{Lat}}_4$ measured on M0 is fit to a constant.]{ \includegraphics[width=0.48\linewidth]{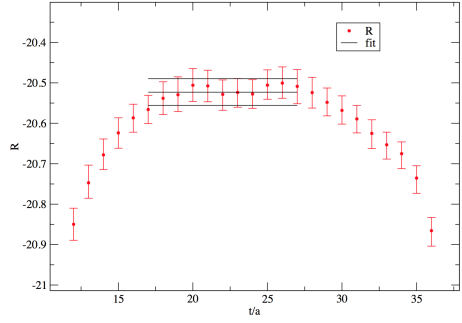} }
	\begin{center}
	\caption{Examples of correlator fits of $B^{\textrm{Lat}}_i$ and $R^{\textrm{Lat}}_i$}
	\end{center}
	\label{fig:corrfits}
\end{figure}

 %---------------------------------------------------------------------------------------
\section{Non-Perturbative Renormalization}\label{sec-3}

The bare parameters are renormalised to ensure a well defined continuum limit and remove any divergences. We use the non-perturbative Rome-Southampton method \cite{Martinelli:1994ty} with non-exceptional kinematics (RI-SMOM) \cite{Sturm:2009kb}.  The RI-SMOM scheme for the SM four-quark operator is described in \cite{Aoki:2010pe}, and the extension to the full SUSY basis in \cite{Boyle:2017skn}.
The renormalised matrix elements can be expressed as,
\begin{equation}
\braket{Q_i}^{\textrm{RI}}(\mu,a) = Z_{ij}^{\textrm{RI}}(\mu,a) \braket{Q_j}^{\textrm{bare}}(a),
\end{equation}
where $Z_{ij}^{\textrm{RI}}(\mu,a)$ denotes the renormalisation factor which, if chiral symmetry breaking effects can be neglected, has block diagonal structure.

\begin{figure}[h]
	\begin{center}
		\begin{fmffile}{renorm} 
				\begin{fmfgraph*}(70,60) 
					\fmfleft{i1,i2} 
					\fmfright{o1,o2} 
					\fmflabel{$d$}{i2} 
					\fmflabel{$d$}{o1} 
					\fmflabel{$s$}{i1} 
					\fmflabel{$s$}{o2} 
					\fmf{fermion,label=$p_2$,label.dist=0.1cm,label.side=left}{i1,v} 
					\fmf{fermion,label=$-p_1$,label.dist=0.1cm,label.side=left}{v,i2} 
					\fmf{fermion,label=$-p_1$,label.dist=0.1cm,label.side=left}{v,o1} 
					\fmf{fermion,label=$p_2$,label.dist=0.1cm,label.side=left}{o2,v}
					\fmfdot{v} 
				\end{fmfgraph*} 
		\end{fmffile} 
	\end{center}
	\caption{The four-quark operator vertex function}%
	\label{fig:vertex}
\end{figure}
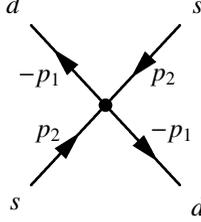

In this method, we require that the projection of the renormalised amputated vertex function (Figure \ref{fig:vertex}) $\Pi^{\textrm{ren}}_i$ is equal to its tree level value. This equation defines $Z^{RI}_{ij}$.

The choice of projector is not unique, in \cite{Garron:2016mva}  we have used two schemes called ($\gamma_\mu, \gamma_\mu$) and ($\slashed{q},\slashed{q}$), which are also the schemes considered in this work. Details on the renormalisation procedure and the definition of these projectors can be found in \cite{Boyle:2017skn}.

\begin{equation}
P_k[\Pi^{\textrm{RI}}_i(p_1,p_2)]_{p^2=\mu^2} = P_k\bigg[\frac{Z_{ij}^{\textrm{RI}}(\mu,a)}{Z_q(\mu,a)}\Pi_j^{\textrm{bare}}(a,p_1,p_2)\bigg]_{p^2=\mu^2} = P_k[\Pi_i^{(0)}]
\end{equation}
Here we present only results obtained through the $(\gamma_\mu,\gamma_\mu)$ projection scheme.

Since the discretisation used in C0(M0) and C1/2(M1/2) only differ by the approximation of the sign function in the infinite Ls limit we have assumed it is valid to reuse the renormalisation calculated for C1/2(M1/2) for C0(M0) for the time being. The renormalisation is calculated on a small subsett of configurations to the matrix element measurements, therefore we propagate the errors on the renormalisation by generating bootstraps according to a gaussian distribution with width equal to the error.

%----------------------------------------------------------------------------------------
\section{Extrapolation to the Physical Point and Continuum Limit}\label{sec-4}

The renormalised parameters are extrapolated to the physical pion mass and continuum limit in an uncorrelated global fit. We use the following two ansatz:

\begin{enumerate}
	\item A chiral and continuum extrapolation of $R_i$ and $B_i$ according to a fit ansatz linear in both $a^2$ and $m_P^2/f_P^2$.
	\begin{equation}
	Y\bigg(a^2,\frac{m^2_{P}}{16\pi^2f_P^2}\bigg) = Y\bigg(0,\frac{m^2_{\pi}}{16\pi^2f_{\pi}^2}\bigg)\bigg[1 + \alpha a^2 + \beta \frac{m^2_{P}}{16\pi^2f_P^2} \bigg]
	\end{equation}
	\item A global fit following NLO SU(2) chiral perturbation theory to a fit function shown below.
	\begin{equation}
	Y\bigg(a^2,\frac{m^2_{P}}{16\pi^2f_P^2}\bigg) = Y\bigg(0,\frac{m^2_{\pi}}{16\pi^2f_{\pi}^2}\bigg)\bigg[1 + \alpha a^2 + \frac{m^2_{P}}{16\pi^2f_P^2}  \bigg(\beta + C_i \log\bigg(\frac{m^2_P}{\Lambda^2}\bigg) \bigg) \bigg]
	\end{equation}
	$C_i$ are the chiral logarithm factors, for $R_i$ we have $C_i = (3/2,3/2,5/2,5/2)$ and for $B_i$ then $C_i$=(-1/2,-1/2,1/2,1/2). $\Lambda$ is the QCD scale.
\end{enumerate}

 We expect the dominant lattice artefacts to be linear in $a^2$ as we use domain-wall fermions. These two methods are equivalent up to the chiral logarithm term, and the difference can indicate how strong the chiral effects from including non-physical pion masses are. 
 The lattice spacings were calculated in \cite{Blum:2014tka} from many of the same ensembles as in this work, therefore a correlation between the data in our global fit is present.. However, in order to decouple this work from the previous work we perform an uncorrelated fit. We propagate the error on lattice spacings by generating bootstraps according to a gaussian distribution with width equal to the error on $a$. The error on the lattice spacings is small (of order 0.5\%) and the contribution of the lattice spacing to the correction of the data is of order 10\% so overall we expect the effect of neglecting these correlations to be small and we believe this approach is justified. 
 When calculating $\chi^2$ in the global fit, we consider the data's deviation from the model in y axis only. The gradient of the slope we obtain in $m_\pi/(4\pi f_\pi)^2$ is small therefore the change in $\chi^2$, were we to instead consider the smallest approach to the fit line, would be negligible.

\section{Results}\label{sec-5}

Plots of the global fit, with the linear fit ansatz, are shown for the ratio parameters in Figure 4. In table \ref{tab:results} we present preliminary results from the global fit for the ratio parameters for both fit ansatz. 
We can see that the fits favour the linear ansatz over the chiral one. The ratio parameter linear ansatz fits are all good fits with $\chi^2$ per d.o.f of less than 1. 
The bag parameters have been measured and renormalised but the global fits have not yet been finalised. Full results including the BSM bag parameters will be included in a future publication.

\begin{table}[h]
	\centering
	\caption{ Preliminary results of the global fit results for both ansatz are presented here alongside the $\chi^2$ per degree of freedom. Results from the previous RBC-UKQCD BSM kaon mixing calculations \cite{Garron:2016mva} are also presented for comparison. All results are presented in the intermediate RI-SMOM$^{(\gamma_\mu,\gamma_\mu)} $scheme at 3GeV}
	\label{tab:results}
	\begin{tabular}{ c |  c c | c c  | c }
		\hline \hline
		& linear fit & $\chi^2$ /dof & chiral PT fit & $\chi^2$ /dof & RBC/UKQCD16 \\
		\hline 
	$R_2$ & -18.69(11) & 0.4 &-18.83(11) & 2.8 & -19.11(43)(31)\\
	$R_3$ & 5.612(41) & 0.7 &5.665(41) & 2.9 & 5.76(14)(16)\\
	$R_4$ & 38.91(21) & 0.3 &39.54(22) & 4.6  & 40.12(82)(188)\\
	$R_5$ & 10.91(6) & 0.3 &11.079(58) & 5.0  & 11.13(21)(83)\\
	\hline \hline
	\end{tabular}
\end{table}

\begin{figure}[h!]
	\label{fig:ratioresults}
	\subfigure{
		\includegraphics[width=0.5\linewidth]{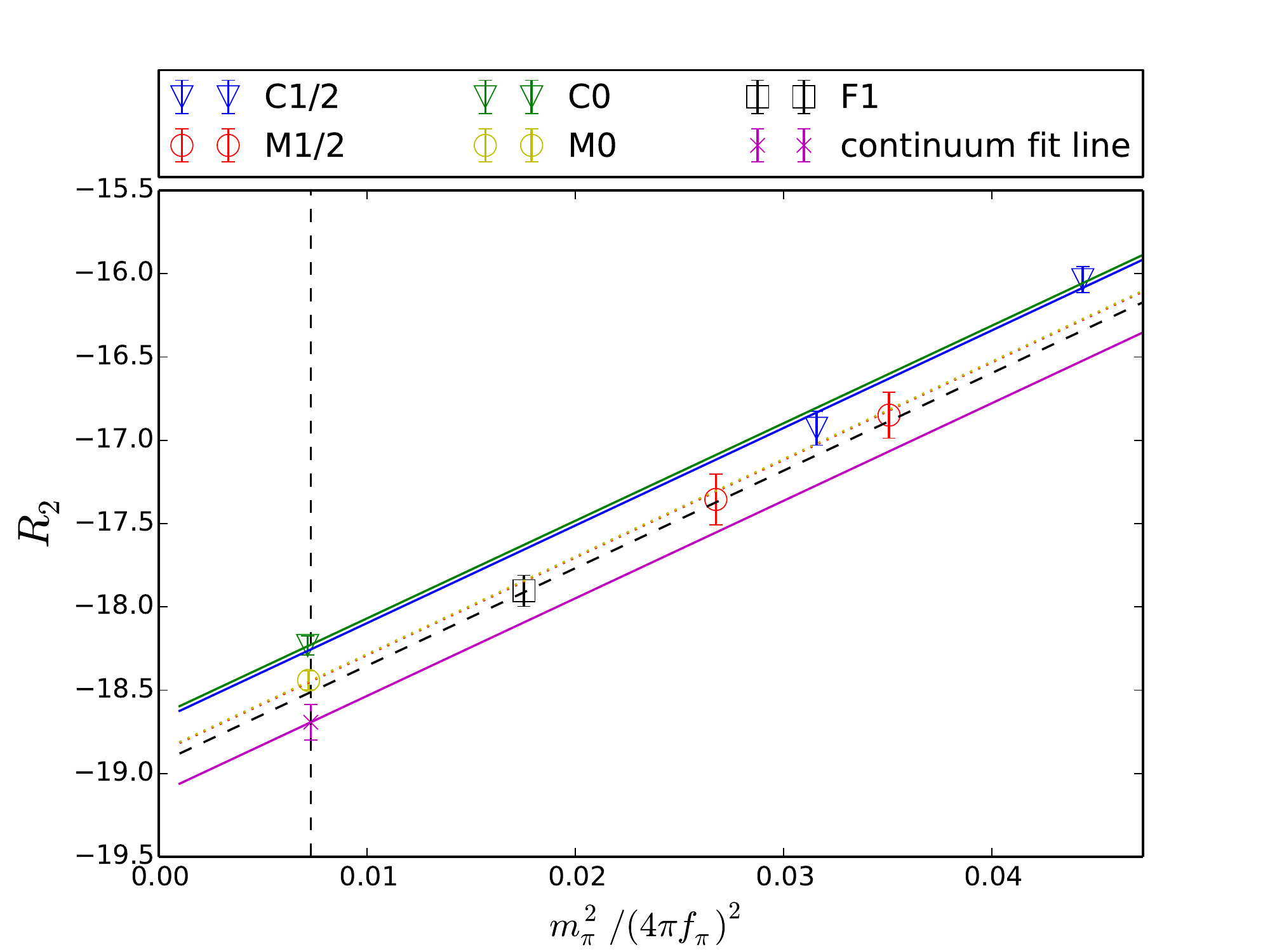}
	}
	\subfigure{
		\includegraphics[width=0.5\linewidth]{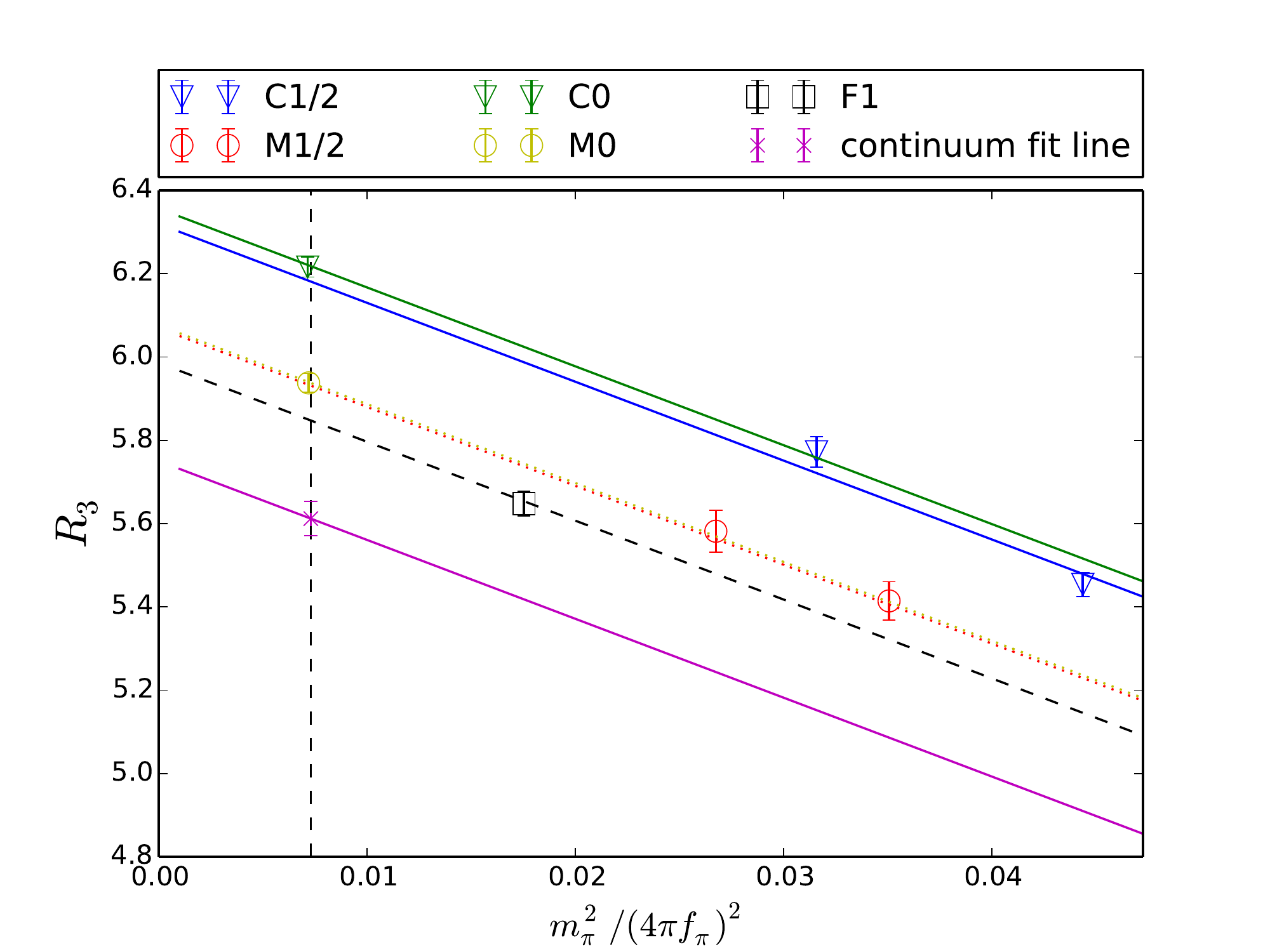}
	}
	\subfigure{
		\includegraphics[width=0.5\linewidth]{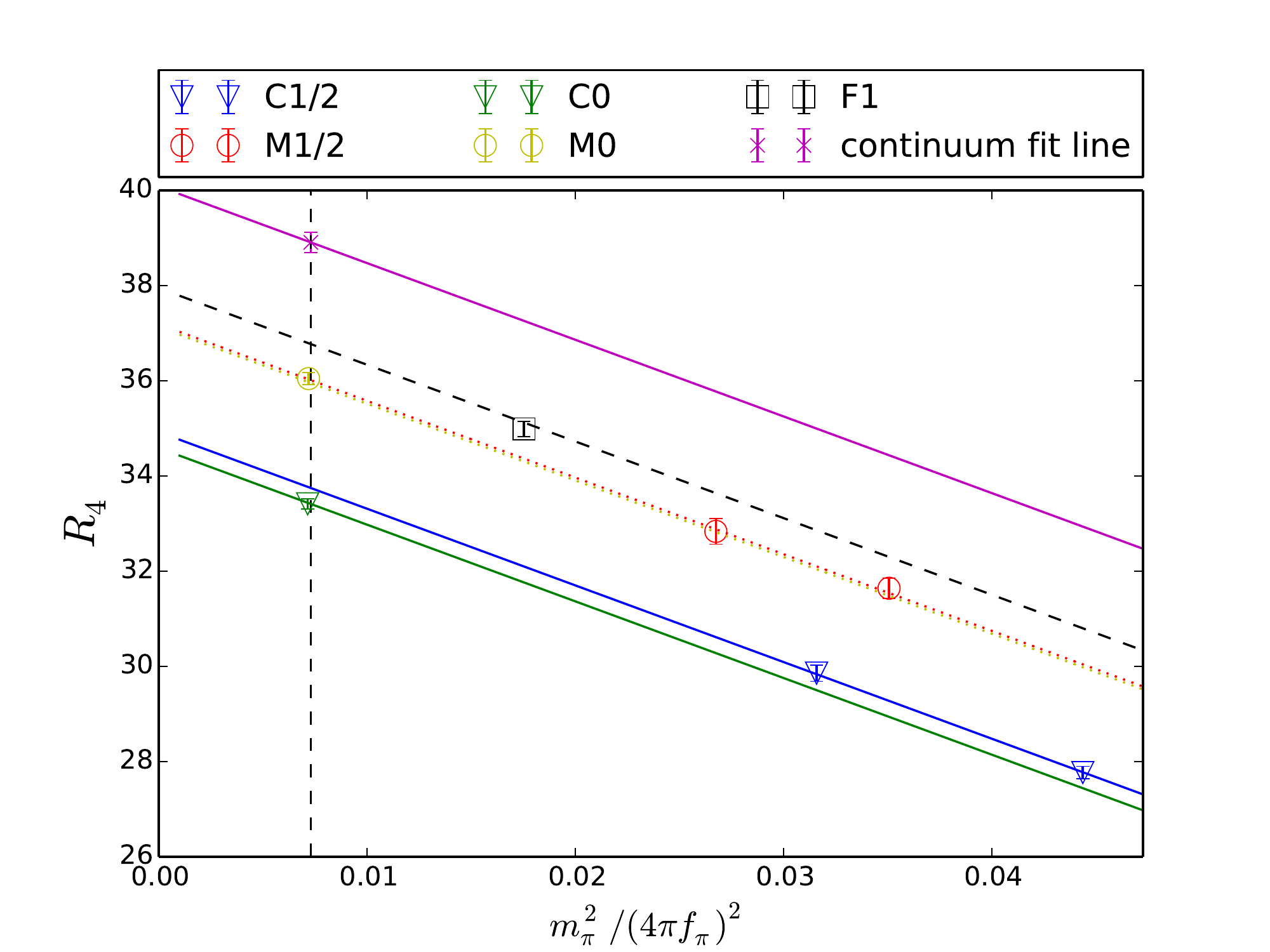}
	}
	\subfigure{
		\includegraphics[width=0.5\linewidth]{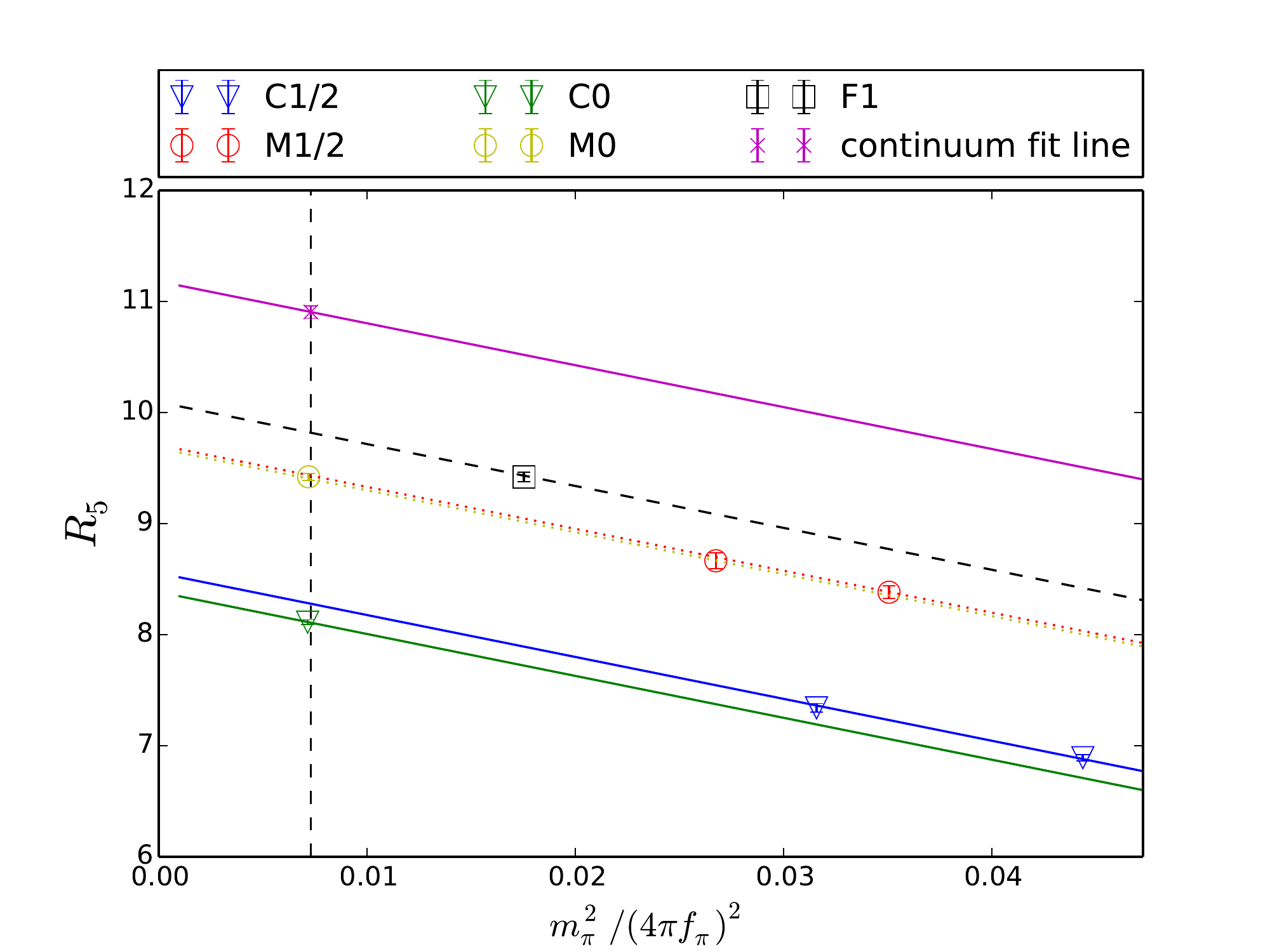}
	}
	\begin{center}
		\caption{The preliminary linear ansatz global fit results of the ratio parameters are presented here. These are shown in the intermediate RI-SMOM$^{(\gamma_\mu,\gamma_\mu)}$ at 3 GeV. The magenta line indicates the continuum limit, with the cross indicating the physical point result. }
	\end{center}

\end{figure}

\section{Conclusion}

We have calculated kaon mixing bag and ratio parameters using $n_f=2+1$ DWF QCD at 3 lattice spacings and several pion masses, including the physical pion mass. We've obtained preliminary results, via a simulataneous chiral/continuum fit, consistent with RBC-UKQCD's previous work \cite{Garron:2016mva} but with statistical errors reduced by a factor of at least 3 for all the ratio parameters. 

We have not yet calculated the sytematic errors but by including measurements at the physical point we have eliminated the systematic error from the chiral extrapolation and the inclusion of a third lattice spacing helps control the continuum extrapolation. Therefore we would expect to have a reduced systematic error too.

 We are in the process of cross-checking the bag parameter fits and are still to convert a renormalisation factor for F1 to $\overline{\text{MS}}$ . Once complete and once the systematic errors have been finalised we will present the full final results in a forthcoming journal publication.

\section{Acknowledgements}
We thank our colleagues in RBC and UKQCD for their contributions and helpful discussions. The measurements in this work were computed on the STFC funded DiRAC facility (grants ST/K005790/1, ST/K005804/1, ST/K000411/1, ST/H008845/1). This research has received funding from the SUPA student prize scheme, Edinburgh Global Research Scholarship, Royal Society Wolfson Research Merit Award WM160035 and STFC (grant ST/L000458/1, ST/M006530/1 and an STFC studentship.) N.G. is supported by the Leverhulme Research grant RPG-2014-118.

\clearpage
\bibliography{lattice2017}

%%%%%%%%%%%%%%%%%%%%%%%%%%%%%%%%%%%%%%%%%%%%%%%%%%%%%%%%%%%%%%%%%%%%%%%%%%%%%
\end{document}